\documentclass[fleqn, usenatbib, useAMS]{mnras}  

\usepackage{graphicx}
\usepackage{amsmath}
\usepackage{amssymb}
\usepackage{multicol}
\usepackage{natbib}

\usepackage[T1]{fontenc}
\usepackage{ae,aecompl}

\usepackage{color}

\newcommand{\beq}[1]{\begin{equation}\label{#1}}
\newcommand{\eeq}{\end{equation}}
\newcommand{\bnq}{\[}
\newcommand{\enq}{\]}

\newcommand{\bea}[1]{\begingroup\setlength\arraycolsep{1.4pt}\begin{eqnarray}\label{#1}}
\newcommand{\eea}{\end{eqnarray}\endgroup}
\newcommand{\bna}{\begingroup\setlength\arraycolsep{1.4pt}\begin{eqnarray*}}
\newcommand{\ena}{\end{eqnarray*}\endgroup}

\newcommand{\equ}[1]{{\eqref{#1}}}

\newcommand{\rcl}{}

\title[Gravity field and rotation of 67P]{Gravity field and solar component of
the precession rate and nutation coefficients of Comet 67P/Churyumov--Gerasimenko}

\author[C. Lhotka, S. Reimond, J. Souchay, O. Baur]{
        C. Lhotka$^{1}$\thanks{Email: christoph.lhotka@oeaw.ac.at},%
        S. Reimond$^{1}$, 
        J.Souchay$^{2}$, 
 	O.Baur$^{1}$
       \\
        $^{1}$Space Research Institute, Austrian Academy of Sciences,
        Schmiedlstrasse, 6, 8042 Graz, Austria,
        \\
        $^{2}${\rcl SYRTE, Observatoire de Paris, PSL Research University, CNRS, 
	Sorbonne Universit\'es,}\\
        {\rcl UPMC Univ. Paris 06, LNE, 61 avenue de l'Observatoire,
	75014 Paris, France.}}

\date{Last updated 2015 Jul 1}

\pubyear{2015}

\begin{document}

\label{firstpage}
\pagerange{\pageref{firstpage}--\pageref{lastpage}}
\maketitle

\begin{abstract}
{The aim of this study is first to determine the gravity field of the comet
67P/Churyumov--Gerasimenko and second to derive the solar component of
the precession rate and nutation coefficients of the spin axis of the comet
nucleus, i.e. without the direct, usually larger, effect of outgassing.} 
{The gravity field, and related moments of inertia, are obtained from 
two polyhedra, that are provided by the OSIRIS and NAVCAM experiments on Rosetta,
and are based on the assumption of uniform density for the comet nucleus.
We also calculate the forced precession rate as well as the nutation coefficients 
on the basis of Kinoshita's theory of rotation of the rigid Earth and adapted it 
to be able to indirectly include the effect of outgassing on the rotational parameters.}
{
The 2nd degree denormalized Stokes coefficients of comet 67P/C-G turn out to be
{\rcl (bracketed numbers refer to second shape model)}
$C_{20}\simeq-6.74 \ [-7.93]\times10^{-2}$, $C_{22}\simeq2.60 \ [2.71]\times10^{-2}$ 
consistent with normalized principal
moments of inertia $A/MR^2\simeq0.13 \ [0.11]$,
$B/MR^2\simeq0.23 \ [0.22]$, with polar moment $c=C/MR^2\simeq0.25$, depending
on the choice of the polyhedron model. The
obliquity between the rotation axis and the mean orbit normal is
$\varepsilon\simeq52^o$, and the precession rate only due to solar torques
becomes $\dot\psi\in[20,30] \ '' / y$. Oscillations in longitude caused by
the gravitational pull of the Sun turn out to be of the order of
$\Delta\psi\simeq1 \ '$, oscillations in obliquity can be estimated to be of
the order of $\Delta\varepsilon\simeq0.5 \ '$.
}  
{
}
\end{abstract}

\begin{keywords}
comets: 67P/Churyumov--Gerasimenko -- celestial mechanics -- gravity field -- 
forced rotational state
\end{keywords}
\maketitle

\section{Introduction}

67P/Churyumov-Gerasimenko (hereafter referred to as 67P in the sequel) is the
target of the ESA space mission Rosetta launched on 2 March 2004. This mission
provided the opportunity for the first safe touchdown of a lander (Philae) on
the surface of a comet nucleus on 12 November 2014. 

The Rosetta data allow determining in a precise way physical and dynamical
characteristics of the comet. This is the purpose of this paper.  First, we
derive a gravity field solution starting from a shape model that is based on
very precise measurements from the Rosetta mission. Second, we
investigate the influence of the gravitational pull of the Sun on the rotation
of comet 67P, i.e. we provide the solar component of the precession rate and
nutation coefficients of the comet's spin axis. We remark that the
gravitational interaction with the other planets (e.g. Jupiter), and
non-gravitational forces generate additional torques that may become orders of
magnitudes larger in comparison with solar torques. While we neglect close
encounters with the planets \citep[see][for the effect of close encounters of
asteroids with the Earth]{SouEtAl14}, outgassing induced effects are important
to understand the long-term evolution in time of the rotational parameters. For
the general inclusion of non-gravitational torques in rotational cometary
dynamics, see e.g. \citet{Sid2008, NeiSidSch2003, NeiSchSid2002, Mys2006,
Mys2007}. In case of 67P the effect of outgassing  has been thoroughly
investigated in the pre-era of the Rosetta mission in \citet{Gut05}. The
authors used different shape models and activity patterns to quantify the
effect for comet 67P and find typical shifts in the spin period of about
$0.1-0.8h$ with typical rates of change of about $0.001-0.05h/d$ (hours per
day).  Furthermore, in \citet{Mot14} the hypothesis has been made that the
rotation rate of 67P may have changed due to cometary activity
during its last perihelion passage.  Variations to the rotational period due
to outgassing are generally accompanied by associated changes in the magnitude
and direction of the angular momentum.  In \citet{Gut05} these changes have
been estimated to be of the order of $10^o$, and at perihelion the angular rate
of change corresponding to the motion of the angular momentum vector amounts to
about $0.01-0.1^o/d$ (degrees per day).  The net torque on the rotation
strongly depends on the water production rate variation over the surface of 67P
that depends itself on the heliocentric distance and insolation conditions of
the comet \citep[see, e.g.][]{KelEtAl15}. In this work the authors describe the
necessity of the accurate modeling of the non-gravitational forces on the basis
of a sublimation model that determines the gas production rate and temperature,
hence the instantaneous force acting on each facet of the real shape model.
Therefore, the realistic modeling of the rotation of comets turns out to be a
challenging problem of great complexity. The full modeling of the rotation of
comet 67P is out of the scope of the present study. However, we aim to
demonstrate the importance of the additional torque on cometary rotation due to
the gravitational pull of the Sun. In the more active phase of the comet
the solar torques are much smaller when compared to outgassing torques, and the
effect on the rotation of additional uncertainties associated with the
outgassing process is usually much larger than solar induced rotational changes
of the comet nucleus. In other words, rotational changes due to solar torques
are in the noise when compared with rotational changes and corresponding
uncertainties associated with outgassing. However, comets usually spend most of
their time in a less active phase of the comet nucleus. Moreover, while solar
torques are continuously acting on the rotation, torques due to outgassing are
only present for some certain amount of time within one orbital period of the
comet.  At the present state of knowledge the actual efficiency of these kinds
of torques is rather unknown. We therefore claim that the accurate
interpretation of the observations of the rotational state of the comet require
the perfect knowledge of all torques acting on the rotation over time. Since
the Rosetta mission allows the accurate determination of the cometary rotation
for the first time it will therefore also allow us to obtain a better insight
into these kinds of torques of different origins. Notice that a deterministic
rigid-body rotation model for any celestial body (planet, moon, asteroid, or
comet) generally serves as a necessary basis for more sophisticated models
including non-gravitational torques. It is the purpose of this paper to provide
this fundamental rotation model.

Our study requires the accurate determination of the principal moments of
inertia in a suitable body-fixed reference frame, that can be derived from the
knowledge of the low-degree gravity field harmonics of comet 67P. Both the
gravity field coefficients and principal moments of inertia of 67P are
currently unknown. For this reason we also develop a new gravity field solution
on the basis of a recent shape model and the assumption of constant mean bulk
density of the comet nucleus.

There is no obvious reason that the density of comet 67P must
be uniform. However, little is known about the internal structure
and density of cometary nuclei, in general: {\it ``Indirect evidences available so
far are not compelling and these questions essentially remain a
matter of speculation ''} \citep[see][]{Lam2015}. The CONSERT radio 
experiment on the Rosetta spacecraft is due to probe the interior
of a comet, i.e. comet 67P, for the first time. No sign of complex
interior structure could be revealed so far. On this basis we think, 
that our uniform density assumption is still the only 
reasonable one.

We remark that the comparison of our gravity field solution with the
gravity field solution obtained from spacecraft orbits of the Rosetta orbiter
will indirectly allow to validate the constant density assumption. A similar
approach has already been applied in the case of asteroid Eros
\citep[see][]{Kon2002}, where the comparison of the 'real' gravity field
with the gravity field obtained from a constant density assumption shows a 
nearly homogeneous asteroid despite its irregular shape.

A statistical analysis of the obliquities, precession rates, and nutation
coefficients for a set of 100 asteroids has been performed by
\citet{LhoEtal2013}. Moreover the determination of these fundamental rotational
parameters for 5 asteroids, that have been targets for past space missions, has
been done by \cite{Pet14}. These studies are based on a theory of rigid body
dynamics constructed by \citet{Kin77} and were implemented for an asteroid in
the case of Eros in \citet{Sou03,Sou05}. The precise modeling and knowledge of
the rotational state of celestial bodies allows investigating important
physical properties of these objects, in particular, mass \& moments of inertia
that are related to their composition and internal structure. Moreover, the
effect of space-weathering on asteroids and comets cannot be satisfactorily
understood without the precise knowledge of the long-term rotational
evolution of theses celestial bodies. \\

In the pre-era of the Rosetta mission the shape and rotational state of the
comet nucleus have been investigated in detail, e.g. in \citet{Lam07, Low12}. A
first analysis of the OSIRIS observations can be found in \citet{Mot14}. First
basic characteristics of the rotation of 67P have already
been identified by the Rosetta mission, like the spin-direction, and precise
rotation period. The comparison of older observations shows that the spin period
decreased by about 0.36 hours since (or during) the perihelion passage in 2009
\citep{Sie15}. It is important to notice that the spin rate is an important
parameter.  An increase of the spin rate may induce more cracks in the
structure of 67P as they have already been identified in the Anuket region
\citep{Rot15} a first indication that the comet may break up into two pieces in
the near future. The Rosetta mission also shows that the nucleus of 67P rotates
about the maximum principal axis of inertia, and the longest axis is nearly
perpendicular to the axes of the individual lobes of the dual lobed comet.
Moreover, the axis with smallest moment of inertia is consistent with being in the
equatorial plane \citep{Sie15}. The coincidence of the axis of rotation with
the axis of maximum moment of inertia suggests that the comet is composed by
weakly bonded icy dust aggregates, with porosity being dominant at small scales
\citep{Sie15}.  Most interestingly, the large obliquity of $52^o$ between the
rotational axis and the axis normal to the orbital plane leads to a greater
exposure to space-weathering of one of the two hemispheres, implying that the
surface structure on the two hemispheres evolved differently over time. The
internal structure of 67P is currently unknown although some hypothesis have
been made on the basis of the relatively low density in comparison to mass and
volume of 67P \citep[e.g., the composition of boulders and rubble
pile,][]{Wei15}. The rotational state of 67P seems also to play a role on
cometary activity since dominant features of the coma's origin have been found
close to the rotational pole exposed by the Sun \citep{Tub15} while the energy
input from the Sun has shown to be smaller in the neck region close to the pole
in comparison with the two lobes \citep{Sie15}. The further investigation of
the complex rotational state of 67P may help to interpret these kinds of
observations. \\

The paper is organized as follows: we derive a new gravity field solution based
on the shape model, called 67P/C-G \citep{ESAwww} in Sec.~\ref{s:shape}. We calculate
mean orbital parameters for 67P based on least square methods in
Sec.~\ref{s:orbit}, and investigate the complex rotation of the comet's nucleus
in Sec.~\ref{s:rotation}.  Results are summarized in Sec.~\ref{s:summary}, and
a discussion about them can be found at the end of the paper.

\section{Gravity field, and moments of inertia of 67P}
\label{s:shape}

\begin{table}
\caption{Physical parameters of 67/P Churyumov-Gerasimenko
used in our study.}
\label{tab:phypar}
\centering
\begin{tabular}{@{}lll@{}}
\hline\hline
Physical parameters \\
\hline
rotation period & $12.4043\pm0.0007 \ h$ & \cite{Mot14} \\
spin-axis & $\alpha=69.3\pm0.1^o$ & \cite{Sie15} \\
& $\delta=64.1\pm0.1^o$ & \\
mass & $10^{13} \ kg$ & \cite{ESAwww} \\
volume & $21.4\pm2 \ km^3$ &  \cite{Sie15}  \\
density & $470\pm45 \ kg/m^3$ & \cite{Sie15} \\
\hline
\end{tabular}
\end{table}

We determine the gravity field up to degree 100, and the moments of inertia
along the principal axes of inertia from the polyhedron shape model (referred
to as 67P/C-G) \citep{ESAwww, RSM}.  The shape model consists of 62908 facets, and
is based on images taken from the OSIRIS and NAVCAM cameras on board of the
Rosetta spacecraft.  In a first step, we rescaled the shape model such that the
volume\footnote{Fundamental physical parameters used in this study are
summarized in Table~\ref{tab:phypar}} of the polyhedron is consistent with the
estimate $V=21.4km^3$.  Next, we determined the mass properties and principal
axes of the rescaled model (hereafter SHA) using a method derived in
\cite{Mir96} under the assumption of a constant mean density $\rho=470kg/m^3$.
Next, we translated the origin of SHA to coincide with the center of mass, and
rotate it such that the $z$-axis becomes aligned with the polar axis of
inertia. The corresponding geometrical transformations can be found in Appendix
A. To obtain the expansion of the gravity field in terms of spherical harmonics
we implement a method based on \citet{WS96, Rei15}, and apply it to the case of
67P (see summary in Appendix B). The method requires to choose the radius of a
circumscribed sphere around the polyhedron model with the condition that
Laplace's equation is fulfilled in the exterior of this reference surface. We
take $R=2800m$ that turns out to be the minimal radius from the
center of mass of SHA that covers all facets of the shape model.

\begin{figure}
\includegraphics[width=0.95\linewidth]{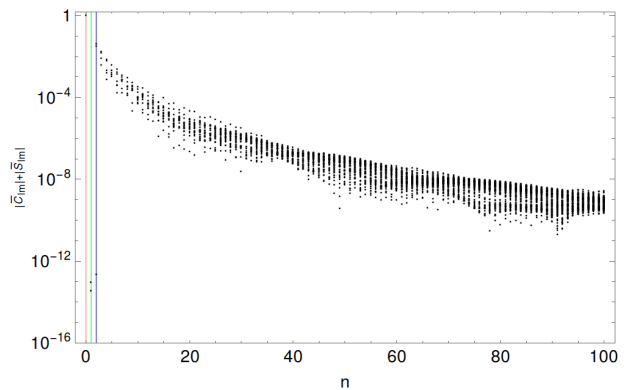}
\caption{Spectrum of the gravity field of 67/P based on
shape model SHA (normalized Stokes coefficients). Zeroth, first,
and second degree harmonics are highlighted in red, green, and blue,
respectively.}
\label{fig:spe}
\end{figure}

The spectrum of the gravity field in terms of normalized coefficients
\citep[see, e.g.][]{Tor12} is shown in Figure~\ref{fig:spe}. We find good
convergence of the series. The low degree, denormalized, coefficients, up to
harmonic degree 3, are summarized in Table~\ref{tab:gra}.  Due to our choice of
a proper coordinate system, we find the coefficients $C_{10}$, $C_{11}$,
$S_{11}$, $C_{21}$, $S_{21}$ and $S_{22}$ to be of the orders of
$10^{-13}\dots10^{-15}$, and the principal coefficients
$C_{20}=-6.74\times10^{-2}$, $C_{22}=2.60\times10^{-2}$. The sole numerical
error in the calculation of the coefficients can be estimated to be of the same
order as $C_{11}$ (that would be zero without numerical errors due to the
proper choice of a suitable body fixed reference system). Since the shape model
67P/C-G comes without any error bars we are unable to determine physical error
bars on the gravity coefficients which could be large because of large errors
on the shape.  To estimate the influence of positional errors in the shape
model on the gravity coefficients given in Table~\ref{tab:gra} we performed the
following test: we first constructed a simplified shape model on the basis of
67P/C-G consisting of 1000 facets only. The simplification has been done by
making use of a quadric edge collapse decimation algorithm~\footnote{ MeshLab
(http://meshlab.sourceforge.net/)} with preservation of boundary, surface
normal and topology of the original mesh. The parameters have been chosen to
allow for offsets of the positions of the vertices within $\pm5\%$ of the
positions of the vertices of the original shape model. These offsets would
therefore correspond to virtual positioning errors of the vertices of the order
of $\pm140m$. Next, we repeated the procedure to determine the gravity field
coefficients on the basis of this simplified shape model and found agreement of
the lower degree Stokes coefficients within $\pm1\%$ of the values published in
Table~\ref{tab:gra}. We notice that knowing the resolution of the camera and
the pointing error, coming from Rosetta orbit miss-modeling, would allow to
estimate the real possibly systematic error. With our numerical experiment we
are able to show that if the real error of the positioning of the vertices of
the shape model is less than $140m$ the shape model induced error on our
results is less than $1\%$.

Recently, a new shape model together with different values for volume $V=18.7km^3$,
and density $\rho=535kg/m3$ have been published \citep[see ][]{Pre2015}. We therefore repeat
our study on the basis of the new shape model, as described above, and find
$C_{20}=-7.93\times10^{-2}$, and $C_{22}=2.71\times10^{-2}$. The difference
($96\%$ agreement in $C_{22}$, but only $85\%$ agreement in $C_{20}$) is consistent
with the different topography of the new shape model: the mass loss due to the thinned 
out part of the new shape model close to the neck region has a much bigger influence
on the mass distribution along the $z$-axis (and therefore also on $C_{20}$) while it 
has a smaller influence on the mass distribution along the $x$-axis (and therefore 
on $C_{22}$). In the following we present our results within the range of possible 
values for the principle gravity harmonics derived on the basis of both shape
models. All values based on the more recent shape model are given in square
brackets.

Using the relations $J_2 M R^2 = C- (A + B)/2$, $C_{22} M R^2 = (B- A)/4$, $C=c
M R^2$ (with $J_2=-C_{20}$) we find the principal moments of inertia $A$, $B$,
$C$ as functions of the normalized polar moment of inertia $c$. We determine
the quantity $c\simeq0.25$ from the requirement that the principal moments of
inertia $A\leq B<C$ (on the diagonal) of the inertia matrix, that can
independently be calculated from \cite{Mir96}, are consistent with the
non-vanishing 2nd degree gravity harmonics that we obtain with the method
developed in \cite{Rei15} (see also Appendix A). Using this value, the other
normalized moments of inertia turn out to be $A/MR^2\simeq0.13 \ [0.11]$ and
$B/MR^2\simeq0.23 \ [0.22]$.  We remark, that the parametric study in $c$ is important
to allow to adapt our results easily once the interior structure of 67P is
known more accurately as it is usually done in planetary studies, e.g. the case
of Mercury \citep[see e.g.][]{Noy13}.

\begin{table}
\caption{Denormalized, low order spherical harmonics up to degree 3 of
shape model SHA. For comparison we provide values obtained 
from a different shape model \citep{Pre2015}, with $V=18.7\pm1.2km^3$ and 
$\rho=535\pm35kg/m^3$ in square brackets.}
\label{tab:gra}
\centering
$
\begin{array}{ccrr}
\hline\hline
\text{Gravity field} \\
\hline
 l & m & C_{lm} & S_{lm} \\
 0 & 0 & 1.0 & 0. \\
 1 & 0 & {\rcl \approx 10^{-13}...}10^{-14} & 0. \\
 1 & 1 & {\rcl \approx 10^{-13}} & {\rcl \approx 10^{-14}} \\
 2 & 0 & -6.74{ \ \rcl [-7.93]}   \times 10^{-2} & 0. \\
 2 & 1 & {\rcl \approx 10^{-13}} & {\rcl \approx 10^{-14}} \\
 2 & 2 & 2.60{\rcl  \ [2.71]}\times 10^{-2} & {\rcl \approx 10^{-14}} \\
 3 & 0 & -2.03{\rcl \ [-1.36]}\times 10^{-2} & 0. \\
 3 & 1 & {\rcl \approx 10^{-4}  \ [10^{-3}]} & {\rcl \approx 10^{-3}} \\
 3 & 2 & {\rcl \approx 10^{-3}} & {\rcl \approx 10^{-3}} \\
 3 & 3 & {\rcl \approx 10^{-4}} & {\rcl \approx 10^{-3}} \\
\hline
\end{array}
$
\end{table}


\section{Mean orbital elements of 67P}
\label{s:orbit}

\begin{table}
\caption{Mean orbital elements derived in this 
study centered around $t_0=2013-01-01$ and obtained within the time window $\pm5y$.}
\label{tab:ele}
\centering
$
\begin{array}{cccccc}
\hline\hline
\text{Orbital parameters} \\
\hline
 a[au] & e & i[{}^o] & \omega[{}^o]  & \Omega[{}^o]  & n[{}^o/d] \\
 3.464 & 0.640 & 7.038 & 12.758 & 50.175 & 0.152 \\
\hline
\end{array}
$
\end{table}

Orbital data for 67/P Churyumov-Gerasimenko is obtained from ephemeris service
\citep{NASADE} based on DE431, and Minor Planet's Data Center \citep{MPC}. In
order to implement the rotation theory in the subsequent section we make use of
mean orbital elements instead of osculating ones. The mean orbital elements are
summarized in Table~\ref{tab:ele}. The osculating ones are obtained in the
following way: we first transform the time series of heliocentric ecliptic
osculating elements from \cite{NASADE} to osculating Keplerian elements using
the mass parameter of the Sun $\mu=2.959\times10^{-5}au^3d^{-2}$. Next, for
each orbital element we fit models of the form:
\beq{eq:fitmod}
X+Y(t-t_0)+Z(t-t_0)^2 \ ,
\eeq  
through each time series, centered around $t_0$=2013-01-01 within the time
window $\pm5y$. Then, from $X$ we immediately obtain the values of mean
semi-major axis $a$, eccentricity $e$, orbital inclination (wrt. to the
ecliptic) $i$, argument of perihelion $\omega$, and longitude of the ascending
node $\Omega$. We validate $X$ from the linear rates of change $Y$ that turn
out to be of the order of  $10^{-6}-10^{-7}$ for $a$, $e$, $i$, and the order
of $10^{-5}$ for $\omega$, and $\Omega$.  The mean motion $n$ can be directly
obtained from constant $Y$ in the fit for mean anomaly $M$, whereas we use $Z$
to validate $Y$ that turns out to be of the order of $10^{-8}$. 



\section{Rotational parameters of comet 67P}
\label{s:rotation}


In this section we investigate the effect of solar torques on the
rotation of the nucleus of comet 67P by taking into account the possible 
variations of the spin period and direction of angular momentum due
to non-gravitational effects \citep[][]{Gut05, KelEtAl15}. Precisely, we 
calculate the secular component characterizing the precession, and its 
short-period oscillations characterizing the nutation of the spin axis.

We assume here in first approximation that 67P can be assimilated to a rigid
body of ellipsoidal shape by means of moments of inertia $A\leq B<C$, with
respect to the semi-major axes of the ellipsoid $\alpha'>\beta'>\gamma'$.
Moreover, as demonstrated by \cite{Sou05} in the case of the asteroid 433 Eros,
we consider that the effect of the triaxial shape of 67P on the rotation is
small enough to be neglected for the purpose of determination of precession
rates and nutation coefficients. The influence of higher degree harmonics (e.g.
$C_{30}$ in Table~\ref{tab:gra}), and non-rigid effects on the rotation are
subject of a follow-on study. 

We also assume that the rotation axis is very close to the figure axis such
that 67P is considered to be in a short axis mode.  This has been confirmed by
the Rosetta mission \citep[][]{Sie15}.  Indeed this mode characterizes the very
large majority of small bodies of the solar system such as asteroids and comets and
represents the natural state resulting from dissipative processes occurring at
relatively short time scales in their history \citep{Bur73}. In this context we
only take into account in our computations the gravitational effect of the Sun
on the rotation of the comet.

In the following, after determination of the obliquity with respect to the mean
orbit of 67P, we express and calculate the precession rate as well as the
nutations in longitude and in obliquity of the comet in the range
of parameters that include possible offsets due to outgassing effects. 

\subsection{Obliquity}

The obliquity $\varepsilon$ can be obtained from the simple equation:
\beq{e:obl}
\cos \varepsilon =\vec{o}\cdot\vec{f} \ .
\eeq
Here, the rectangular coordinates of the spin-pole $\vec{f}$  are $\vec{f} =
(\cos\beta\cos\lambda, \ \cos\beta\sin\lambda, \ \sin\beta)$, where $\lambda$
and $ \beta$ stand respectively for the ecliptic longitude and latitude of
$\vec{f}$.  The rectangular coordinates of the orbit-pole $\vec{o}$ are
themselves directly determined from the inclination $i$ and the longitude of
the node $\Omega$ of the orbit: $\vec{o} =  (\sin i\sin\Omega, \
-\cos\Omega\sin i, \  \cos i)$. As the spin-pole is given by its equatorial
coordinates $(\alpha, \delta)$ whereas we are doing our calculations in an
ecliptic frame, we make use of the transformation from equatorial to ecliptic
coordinates
\bnq
\left(
\begin{array}{c}
 q_1 \\
 q_2 \\
 q_3 \\
\end{array}
\right)=
R_1\left(-\varepsilon_E\right)
\left(
\begin{array}{c}
 \cos \alpha  \cos \delta  \\
 \sin \alpha  \cos \delta  \\
 \sin \delta  \\
\end{array}
\right) \ ,
\enq
with rotation matrix $R_1$ around the $x$-axis. 
Here $\varepsilon_E$ represents the nominal value of the Earth obliquity.
Thus we calculate $(\lambda, \beta)$ from
\bnq
\lambda=\tan^{-1}\left(\frac{q_2}{q_1}\right) \ , \quad \beta=\sin^{-1}q_3 \ .
\enq 
Using $\varepsilon_E=23^o26'21.448''$ (J2000) and  $(\alpha,
\delta)\simeq(69.3^o,64.1^o)$ we find $(\lambda,\beta)=(78^o,42^o)$, and
therefore we find the obliquity of 67P to be $\varepsilon\simeq52^o.062$, in
agreement with the published value of $52^o$ in \cite{ESAwww2, Sie15}.
In the following we allow for an offset $\delta\varepsilon$ of $10^o$ in $\varepsilon$ to
include possible variations of the spin-axis due to changes in the direction of
the angular momentum vector during the rendezvous of the Rosetta mission
\citep[][]{Gut05}.

\subsection{Determination of the precession rate \& nutation coefficients}

\begin{figure}
\centering
\includegraphics[width=.95\linewidth]{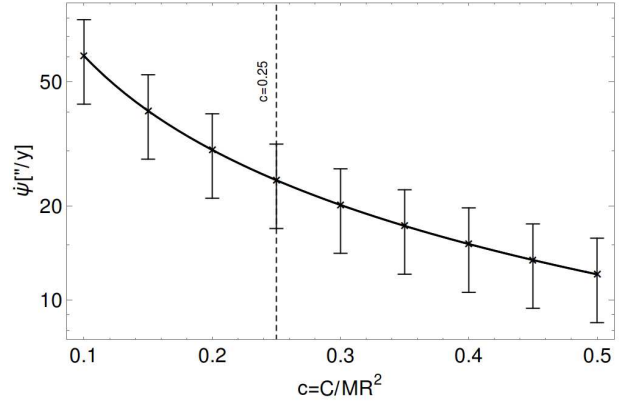}
\caption{Mean constant precession rate of 67/P for $\varepsilon=52\pm10^o$
and $\omega_s=12.4043\pm1h$ for different polar moment of inertia $c$.}
\label{fig:pre}
\end{figure}

The general theoretical framework to model the rotation of a given
celestial body as the comet 67P has been constructed by \cite{Kin77}.
Starting from this framework \cite{Pet14} have proposed formulae,
valid up to order $4$ in $e$, for the precession rate $\dot \psi$, 
the nutation  of the longitude of the node $\Delta\psi$, and the 
oscillations of the obliquity $\Delta\varepsilon$. These formulae 
were successfully applied on celestial bodies with well defined physical constraints, 
as (1) Ceres, (4) Vesta, (433) Eros, (2867) Steins and (25143) Itokawa. Since 
the eccentricity, $e=0.64$, of 67P is much larger than for these kinds of 
objects we need to develop these formulae to much higher order in $e$ to be 
able to apply them for the present case of 67P as well. At this point,
we provide the general formula for the precession rate up to order $16$ 
in $e$, while we only summarize the formulae for $\Delta\psi$, $\Delta\varepsilon$ up 
to 4th order in $e$ (but still use 16th order formulae in our
calculations). The precession rate up to order $16$, according to \citet{Pet14}, is given by:
\bea{e:pre} 
\dot\psi&=&\bigg[
1
+\frac{3}{2}e^2
+\frac{15}{8}e^4
+\frac{35}{16}e^6
+\frac{315}{128}e^8
+\frac{693}{256}e^{10} + \nonumber \\
& &\frac{3003}{1024}e^{12}
+\frac{6435}{2048}e^{14}
+\frac{109395}{32768}e^{16}
\bigg]\times
\frac{K}{2}\cos\varepsilon \ ,
\eea
where the constant $K$ together with its possible ranges during one orbital
period of the comet is given by
\bna
K&=&\frac{3n^2}{\omega_s}\times H_d \simeq \frac{ 8.9 \ [10.5]}{c} \ ''/y \ .
\ena
Here,  $n=0.1529^o/d$ is the mean motion of the comet, and
$\omega_s\in(644.569,757.609)^o/d$ is the spin frequency consistent with the
rotation period $T=12.4043\pm1h$ that includes possible variations
due to outgassing of the order of $\pm 1h$.  The expression
also contains the dynamical ellipticity $H_d$ that is related to the moments of
inertia (and therefore 2nd degree gravitational harmonics) by:
\bna
H_d&=&\frac{2C-A-B}{2C} \simeq \frac{ 0.0674 \ [0.0793]}{c}\ .
\ena
Since the polar moment of inertia may be sensitive to the structure models of
the interior of the comet  - that we do not take into account at the present
time - we provide a parameter study of $\dot\psi$ for different values
of polar moments of inertia $c$ in Figure~\ref{fig:pre}. As we can see the
current precession rate may vary between  $12 \ ''/y$ to $30 \ ''/y$ within the
interval $0.2\leq c\leq0.5$.  For $c=0.25$ the actual value of $\dot\psi$ turns
out to lie in the interval $[20, 30] ''/y$.  This precession rate due to the sole
gravitational forcing of the Sun is comparable to the corresponding precession
rate of the Earth (1/3 of the total lunisolar part) that is to
say $15 \ ''/y$.

\noindent The nutation in longitude $\Delta \psi$ and in obliquity $\Delta
\varepsilon$ can be expressed  starting from the mean anomaly $M$ and its sole
harmonics as follows:
\bea{e:serA}
\Delta \psi = \Bigg[
&&\left(
 C' \left(\frac{e}{2}-\frac{e^3}{12}\right)+\frac{27 e^3}{8}+3 e
\right)
\frac{\sin\left(M \right)}{n} \ +\cr
&&\left(
 C' \left(-\frac{41 e^4}{48}+\frac{5 e^2}{2}-1\right)+\frac{7 e^4}{2}+\frac{9 e^2}{2}
\right)
\frac{\sin\left(2M \right)}{2n} \ +\cr
&&\left(
 C' \left(\frac{123 e^3}{16}-\frac{7 e}{2}\right)+\frac{53 e^3}{8}
\right)
\frac{\sin\left(3M \right)}{3n} \ +\cr
&&\left(
 C' \left(\frac{115 e^4}{6}-\frac{17 e^2}{2}\right)+\frac{77 e^4}{8}
\right)
\frac{\sin\left(4M \right)}{4n} \ -\cr
&&
 \frac{845 C' e^3}{48}
\frac{\sin\left(5M \right)}{5n} \ -
 \frac{533 C' e^4}{16}
\frac{\sin\left(6M \right)}{6n}
\Bigg]\times
\frac{K\cos\left(\varepsilon\right)}{2} \ -\cr
S'\Bigg[
&& \left(\frac{e^3}{24}-\frac{e}{2}\right)
\frac{\cos\left(M \right)}{n} \ +
\left(\frac{37 e^4}{48}-\frac{5 e^2}{2}+1\right)
\frac{\cos\left(2M \right)}{2n} \ + \cr
&& \hskip -.3in \left(\frac{7 e}{2}-\frac{123 e^3}{16}\right)
\frac{\cos\left(3M \right)}{3n} \ +
 \left(\frac{17 e^2}{2}-\frac{115 e^4}{6}\right)
\frac{\cos\left(4M \right)}{4n} \ + \cr
&&
 \frac{845 e^3}{48}
\frac{\cos\left(5M \right)}{5n} \ +
 \frac{533 e^4}{16}
\frac{\cos\left(6M \right)}{6n}
\Bigg]\times
\frac{K\cos\varepsilon}{2} \ ,
\eea
and
\bea{e:serB}
\Delta\varepsilon = \Bigg[
&&
 \left(\frac{e^3}{12}-\frac{e}{2}\right)
\frac{S'\sin\left(M \right)}{n} \ -
\left(\frac{e^3}{24}-\frac{e}{2}\right)
\frac{C'\cos\left(M \right)}{n} \ +\cr
&&
 \left(\frac{41 e^4}{48}-\frac{5 e^2}{2}+1\right)
\frac{S'\sin\left(2M \right)}{2n} \ - \cr
&&
\left(\frac{37 e^4}{48}-\frac{5 e^2}{2}+1\right)
\frac{C'\cos\left(2M \right)}{2n} \ +\cr
&&
 \left(\frac{7 e}{2}-\frac{123 e^3}{16}\right)
\frac{S'\sin\left(3M \right) - C'\cos\left(3M\right)}{3n} \ +\cr
&&
 \left(\frac{17 e^2}{2}-\frac{115 e^4}{6}\right)
\frac{S'\sin\left(4M \right) - C'\cos\left(4M \right)}{4n} \ +\cr
&&
 \frac{845 e^3}{48}
\frac{S'\sin\left(5M \right) - C'\cos\left(5M \right)}{5n} \ +\cr
&&
 \frac{533 e^4}{16}
\frac{S'\sin\left(6M \right) - C'\cos\left(6M\right)}{6n}
\Bigg]\times
\frac{K\sin\varepsilon}{2} \ .
\eea
In these formulae the coefficients $C'$, $S'$ are obtained from $C' = \cos (2
\omega - 2\Lambda)$ and $S' = \sin (2 \omega - 2\Lambda)$.  Here $\Lambda$
stands for the angle along the orbital plane between the equinox $\vec{\Gamma}$
and the ascending node $\vec{N} = (\cos \Omega, \sin \Omega, 0)$ of the orbital
plane with respect to the inertial plane of ecliptic J2000.0, where the  unit
vector  $\vec{\Gamma}$ is given by the following vectorial product:
\beq{e:gamma} 
\vec \Gamma  = {1 \over \sin \varepsilon} \vec{f} \times \vec{o}
\ .  
\eeq 
Notice, that in \eqref{e:serA} and \eqref{e:serB} the orbital
longitude $\lambda$ of the perturbing body, i.e. the Sun, is counted from the
"equinox" $\vec{\Gamma}$ of the comet which is the ascending node of the
relative orbit  of the Sun (as determined from the comet) with respect to the
comet's equatorial plane. Therefore: $\lambda = 180^{\circ} + \omega + \nu -
\Lambda$, where $\nu$ is the true anomaly. The application of the expressions
above leads to: $\Lambda = -57^{\circ}$ and $C'=-0.76114$ and $S'=0.6485$. We
provide the nutation series, up to order $16$ in eccentricity $e$, for the
parameters of Table~\ref{tab:phypar} and $c=0.25$ in Table~\ref{nutA}. We
clearly observe that $\dot M$ is the fundamental period of the nutation motion.

\begin{table}
\caption{Nutation series in $\Delta\psi$ and $\Delta\varepsilon$ for actual
physical parameters of 67P and $c=0.25$:
$c_k$ / $s_k$ amplitudes in $\cos$ / $\sin$ for $\Delta\psi$;
$c_k'$ / $s_k'$ amplitudes in $\cos$ / $\sin$ for $\Delta\varepsilon$.
}
\label{nutA}
\centering
$
\begin{array}{rrrrrl}
\hline\hline
c_k[''] & s_k[''] & c_k'[''] & s_k'[''] & P[y] & arg \\
\hline
 2.2898 & 38.1269 & -3.4469 & -2.7937 & 6.44 & M \\
 -0.3419 & 17.2808 & 0.5148 & 0.5318 & 3.22 & 2 M \\
 -1.4352 & 10.9079 & 2.1605 & 1.9000 & 2.14 & 3 M \\
 -1.8545 & 7.8207 & 2.7917 & 2.4168 & 1.61 & 4 M \\
 -1.9383 & 5.9330 & 2.9178 & 2.5112 & 1.28 & 5 M \\
 -1.8525 & 4.6407 & 2.7887 & 2.3930 & 1.07 & 6 M \\
 -1.6852 & 3.6773 & 2.5369 & 2.1730 & 0.92 & 7 M \\
 -1.4866 & 2.9417 & 2.2378 & 1.9146 & 0.80 & 8 M \\
 -1.3273 & 2.4026 & 1.9981 & 1.7079 & 0.71 & 9 M \\
 -1.1585 & 1.9661 & 1.7440 & 1.4898 & 0.64 & 10 M \\
 -0.4761 & 1.0516 & 0.7168 & 0.6130 & 0.58 & 11 M \\
 -0.2112 & 0.6301 & 0.3180 & 0.2724 & 0.53 & 12 M \\
\hline
\end{array}
$
\end{table}

\begin{figure}
\centering
\includegraphics[width=.95\linewidth]{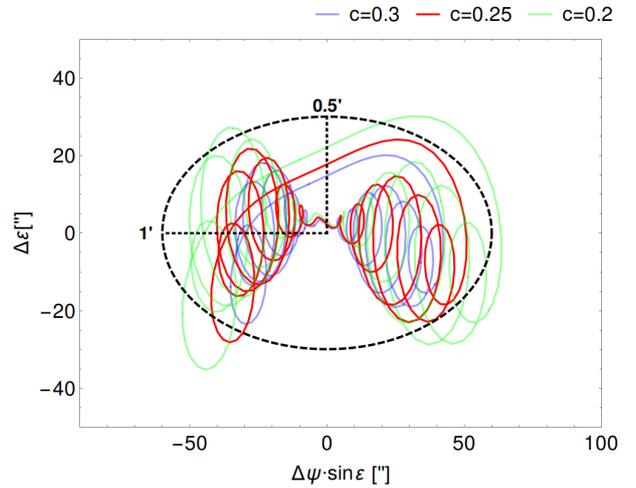}
\caption{Complex bi-dimensional nutational motion
($\Delta\psi\sin\cdot\varepsilon$, $\Delta\varepsilon$) of 67P.  Semi-major
axes of the dashed ellipse correspond to order of magnitudes of 
nutation amplitudes in the $\Delta\psi\cdot\sin\varepsilon$ , $\Delta\varepsilon$ -directions.}
\label{f:bi}
\end{figure}

We also compute the sole nutational part as a function of time resulting in a
bi-dimensional motion ($\Delta\psi \cdot \sin \varepsilon$ ,
$\Delta\varepsilon$) projected to the equatorial plane as shown in
Figure~\ref{f:bi}: the spin axis describes a complex multi-periodic closed loop
whose amplitude varies within $0.5'$ to $1'$ within one orbital period $0\leq
M\leq 360^o$. We also provide, for reference, a solution with $c=0.2$ and
$c=0.3$ and see that the amplitudes (in $\Delta\psi\cdot\sin\varepsilon$, 
$\Delta\varepsilon$) decrease for larger values of the moment of inertia $c$. 
Typical amplitudes of nutation in
$\Delta\psi\cdot\sin\varepsilon$ roughly range within $(-50 \ '', 65 \ '' )$, for
$c=0.2$, from $-35 \ ''\leq \Delta\psi\cdot\sin\varepsilon \leq 45 \ ''$, for 
$c=0.3$, and within $(-40 \ '',  50 \ '')$ for $c=0.25$.  The corresponding amplitudes in
$\Delta \varepsilon$ range from about $-35 \ ''\leq \Delta\varepsilon \leq 30 \
'' $, for $c=0.2$, from  $-25 \ ''\leq \Delta\varepsilon \leq 20 \ '' $, for
$c=0.3$, and from $-30 \ ''\leq \Delta\varepsilon \leq 25 \ ''$ for $c=0.25$.

\begin{figure}
\centering
\includegraphics[width=.95\linewidth]{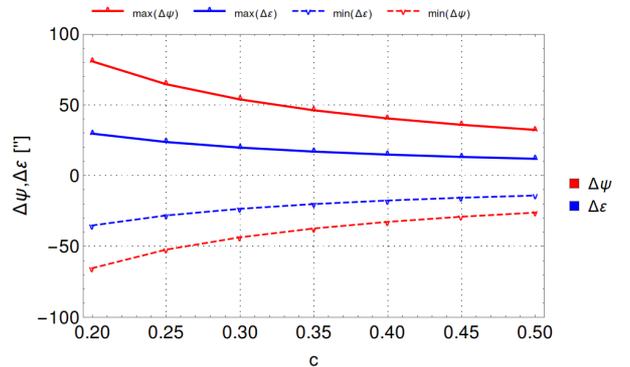}
\caption{Maximum ($mx$) and minimum ($mn$) values of nutation 
of comet 67P, for various polar moments of inertia $0.2\leq c\leq0.5$: 
$mx(\Delta\psi)$ (thick, red),
$mx(\Delta\varepsilon)$ (thick, blue), 
$mn(\Delta\varepsilon)$ (dashed, blue),
$mn(\Delta\psi)$ (dashed, red).}
\label{f:c-study}
\end{figure}

If we repeat our study for various values of $c$ in the interval $(0.2,0.5)$ we
obtain the results summarized in Figure~\ref{f:c-study}, where we show maximum
and minimum values for nutation coefficients defined as follows: we calculate
the time series of the nutation, for different $c$, over one full revolution
period of 67P.  The time series therefore is the superposition of various
trigonometric terms with different periods. Since the different harmonics may
sum up or cancel out each other - depending on the actual value of the mean
anomaly $M$ - we calculate the furthest points along the
$\Delta\psi\cdot\sin\varepsilon$ and $\Delta\varepsilon$ - directions, and
denote by $mx(\Delta\psi\cdot\sin\varepsilon)$, $mx(\Delta\varepsilon)$ the
maximal values into the positive, and by $mn(\Delta\psi\cdot\sin\varepsilon)$,
$mn(\Delta\varepsilon)$ the maximal values into the negative directions.
We notice in particular that amplitudes in oscillations in longitude are 
typically significantly larger than oscillations in obliquity, and that nutation 
coefficients decrease with increasing polar moment of inertia $c$. \\

\subsection{Influence of outgassing-induced effects}

\begin{figure}
\centering
\includegraphics[width=.95\linewidth]{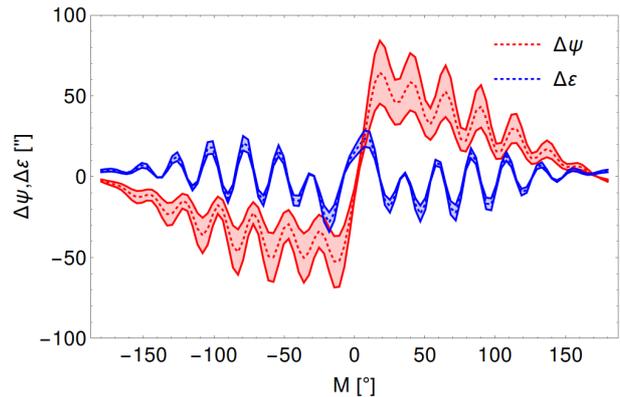}
\caption{Nutation series vs. mean anomaly $M$ for normalized
polar moment of inertia $c=0.25$. Dotted lines: actual
physical parameters of 67P; shaded regions: possible offsets
due to the outgassing effect. }
\label{f:nut-study}
\end{figure}

In this section we investigate the influence of outgassing-induced effects on
the time series of the nutation parameters of comet 67P using the proposed
values for the changes in spin period and direction of angular momentum after
\citet{Gut05, KelEtAl15}. We follow our approach to obtain the results for the
precessional motion and allow offsets $\delta\varepsilon$ of $\pm10^o$ from the
nominal value of the obliquity $\varepsilon$ as well as offsets
$\delta\omega_s$ of $\pm1h$ in the rotation period of the comet. As a
consequence the constants $K$, $C'$, and $S'$ in \equ{e:serA}, \equ{e:serB}
will change in well determined intervals too. To account for the full ranges of
the intervals we therefore look, for each value of $M$, for the minimum and
maximum values of $\Delta\psi$ and $\Delta\varepsilon$, respectively. The
results for $c=0.25$ are shown in Figure~\ref{f:nut-study}.  We observe that
the nutation amplitude in $\Delta\psi$ is about twice as big as the amplitude
in $\Delta\varepsilon$. We also clearly see that the maximal nutation
amplitudes are found close to $M=0$ that corresponds to the time of next
perihelion passage \citep[on August 13, 2015][]{MPC}, while the amplitudes
decrease by orders of magnitude close to aphelion ($M=180^o$). The variations
in obliquity and rotation periods are more present in longitudinal
directions.\\

\section{Conclusions}
\label{s:summary}

In this paper we have determined for the first time the coefficients of the
gravity field of the comet 67P by using a shape model based on very precise
data measurements from the Rosetta mission. Then we have investigated the
motion of its spin-axis due the gravitational forcing of the Sun. We have found
that this motion is rather complex due to both the irregular shape and the high
value of the orbital eccentricity of the comet.  With a value of $c = 0.25$ we
have calculated a precession rate $\dot\psi\in[20,30] \ ''/y$, comparable to the solar
part of the precession of the Earth (roughly $15'' /y$).  Moreover we have
found a maximum amplitude in nutation in longitude of about $\Delta\psi \approx
1 \ '$, and a maximum nutation amplitude in nutation in latitude of about
$\Delta\varepsilon \approx 0.5 \ '$.  We notice that these
nutation amplitudes are much larger than the corresponding ones for the Earth
(respectively roughly $10 ''$ to $20 ''$). As can be seen in
\equ{e:serA}--\equ{e:serB} the reason lies partly in the large eccentricity,
but also in the large value of the dynamical ellipticity $H_d$ of the comet and
consequently of the scaling factor $K$.

In \cite{Sie15} the nucleus structure and activity of comet 67P
have been investigated based on data of the OSIRIS scientific imaging system on board
the Rosetta space-craft. The authors found no obvious evidence for complex
rotation of the comet nucleus and were able to constrain any motion of the spin-axis
to $<0.3^o$ over $55$ days. In this work we predict a complex motion of the
spin axis of 67P over the comet's orbital period of $6.4$ years,
that is in agreement with the bounds given by \cite{Sie15}, namely in terms of the precession
rate and nutation coefficients of the comet's spin axis based on a rigid body
approximation. This preliminary work looks necessary to any further work dealing
with the short or long term evolution of the rotational state of comet 67P, in
particular concerning the variations of its spin axis in space.

We also performed a parametric study in normalized polar moment of inertia
linked to rotational parameters.  The relatively small value of the moment of
inertia factor $c$ is consistent with the thinned out part of 67P along the
spin-axis direction. However, the small value of $c$ may also indicate a
possible differentiated interior structure, and our parametric study should
allow to validate different density profiles. For this purpose, the extended
Rosetta mission period will be crucial to improve the chances to detect the
precession rate and nutation coefficients of comet 67P.

Our study is focussed on the influence of solar torques on the rotational
parameters of the comet. We investigated the interplay between these torques
and the outgassing-induced effects on the basis of recent Rosetta findings
\citep[][]{KelEtAl15}.  With this we are able to provide a better insight
into the sensitivity of the solar component to the specific rotational state of
the comet. Our study may therefore serve as a good starting point for better
models of cometary rotational dynamics.

\vskip.05in
\noindent {\bf Acknowledgments}

We thank an anonymous reviewer for valuable suggestions, and
F. Preusker from DLR for providing us the most accurate shape 
model for comet 67P \citep[][]{Pre2015}.

\bibliographystyle{mnras}
\bibliography{biblio}

\appendix

\section{Geometric transformations}

We implement MatLab and Mathematica programs to rescale the original shape
model 67P/C-G such that the mass, volume, and density are consistent with
Table~\ref{tab:phypar}, and translate its center of mass to the origin. The 
inertia tensor $I$ of the physical shape model is calculated on the basis of
\cite{Mir96} to find the principal axes. Using the eigensystem of $I$ we 
diagonalize the inertia matrix. The center of mass of the rescaled shape model 
before translation turns out to be:
\bnq
(X_0,Y_0,Z_0)=(2.7525\times10^2m, \ 1.0197\times10^2m, \ 0.1093\times10^{2}m) \ .
\enq
The rotation matrix composed by the eigenvectors becomes:
\bnq
PA=
\left(
\begin{array}{ccc}
 -0.9974958 & 0.004987262 & 0.07054984 \\
 -0.02099773 & -0.9734178 & -0.2280722 \\
 0.06753702 & -0.2289824 & 0.9710849 \\
\end{array}
\right) \ .
\enq
The rescaled, translated and rotated shape model SHA is the basis for the
calculation of the gravitational field provided in Table~\ref{tab:gra}. \\

{\bf Remark.} The center of mass and inertia matrix can also be derived
from the spherical harmonic coefficients of degree 1 and 2.
Let $A$, $B$, $C$, $D$, $E$, $F$ be the diagonal \& off-diagonal matrix 
elements of the mass-inertia tensor \citep[see in more detail][]{Tor12}:
\bnq
\begingroup
\renewcommand*{\arraystretch}{1.5}
\begin{array}{ll}
A = \int_V\left(Y^2+Z^2\right)dm & B = \int_V\left(X^2+Z^2\right)dm \ , \nonumber \\
C = \int_V\left(X^2+Y^2\right)dm & D = \int_V Y Z dm \ , \nonumber \\
E = \int_V X Z dm & F = \int_V X Y dm \ .
\end{array}
\endgroup
\enq

\noindent Using the standard definitions $C_{lm}$, $S_{lm}$ with $l\leq2$, $m\leq l$, and of 
the center of mass and inertia matrix can be put into
\bnq
(X_0,Y_0,Z_0)=R(C_{11},S_{11},C_{10}) \ ,
\enq
and
\beq{ITEN}
I=MR^2\cdot\left(e_1,e_2,e_3\right)
\eeq
with
\bnq
e_1=\left(
\begin{array}{l}
 c-C_{10}^2+C_{20}-2 C_{22}-S_{11}^2 \\
 C_{11} S_{11}-2 S_{22} \\
 C_{10} C_{11}-C_{21} \\
\end{array}
\right) \ ,
\enq
\bnq
e_2=\left(
\begin{array}{l}
 C_{11} S_{11}-2 S_{22} \\
 c-C_{10}^2-C_{11}^2+C_{20}+2 C_{22} \\
 C_{10} S_{11}-S_{21} \\
\end{array}
\right) \ ,
\enq
\bnq
e_3=\left(
\begin{array}{l}
 C_{10} C_{11}-C_{21} \\
 C_{10} S_{11}-S_{21} \\
 c-C_{11}^2-S_{11}^2 \\
\end{array}
\right) \ ,
\enq
where we make use of the parametrization $C=c MR^2$. Inserting 
the values of Table~\ref{tab:gra} into \equ{ITEN}, and
equating with $I$ obtained directly with the method proposed in
\cite{Mir96} allows to obtain $c\simeq0.25$.

\section{Determination of the gravity field}

We use the method of least squares adjustment to determine in Table~\ref{tab:gra}
the gravitational field coefficients $C_{lm}$, $S_{lm}$ in the spherical harmonics expansion
of the gravitational potential \citep{Tor12}. Let $\mathbf{l}$ be the vector
of evaluations of the potential, $\mathbf{A}$ be the design matrix, and
$\mathbf{x}$ be the parameter vector of gravity harmonics:
\bna
\label{eq:lsavecs}
\mathbf{l} = \left[ \begin{array}{cccc} U_1 & U_2 & \hdots & U_K   \end{array} \right]^{\sf T}, 
\qquad 
\mathbf{A} = \left[ \begin{array}{cc} \mathbf{A}_C & \mathbf{A}_S  \end{array} \right], 
\qquad 
\mathbf{x} = \left[ \begin{array}{cc} \mathbf{x}_C & \mathbf{x}_S  \end{array} \right]^{\sf T}.
\ena
The gravitational potential values $U_k$ of the shape model SHA, with $k=1,\hdots,K$, 
are computed for evenly distributed points on the surface of the reference sphere
using the algorithm presented in \cite{WS96}. In order to
guarantee for good coverage and highly overdetermined equation systems, a
Reuter grid on the reference sphere of radius $R=2800m$ with $200$
meridional points is used, which yields in total $K=50831$ values in
$\mathbf{l}$.  The design matrix establishes the relation between the
evaluations of the potential and the unknown coefficients of the potential. It
is given by the partial derivatives of the spherical harmonics expansion with
respect to the coefficients $C_{lm}$ ($\mathbf{A}_C$) and $S_{lm}$
($\mathbf{A}_S$):
\bna
\mathbf{A}_C &= \left[ \begin{array}{ccccc}
\partial U_1/\partial C_{0,0} & \partial U_1/\partial C_{1,0} & \ldots & \partial U_1/\partial C_{N,N} \\
\partial U_2/\partial C_{0,0} & \partial U_2/\partial C_{1,0} & \ldots & \partial U_2/\partial C_{N,N} \\
\vdots & \vdots & \vdots & \ddots & \vdots \\
\partial U_K/\partial C_{0,0} & \partial U_K/\partial C_{1,0} & \ldots & \partial U_K/\partial C_{N,N} 
\end{array} \right] \label{eq:lsadesign1}    \nonumber    \\
\mathbf{A}_S &= \left[ \begin{array}{ccccc}
\partial U_1/\partial S_{1,1} & \partial U_1/\partial S_{2,1} & \ldots & \partial U_1/\partial C_{N,N} \\
\partial U_2/\partial S_{1,1} & \partial U_2/\partial S_{2,1} & \ldots & \partial U_2/\partial S_{N,N}  \\
\vdots & \vdots & \vdots & \ddots & \vdots \\
\partial U_K/\partial S_{1,1} & \partial U_K/\partial S_{2,1} & \ldots & \partial U_K/\partial S_{N,N}  
\end{array} \right] \label{eq:lsadesign2} \\
\ena
The coefficients in $\mathbf{x}$ are of course ordered accordingly inside the parameter vector:
\bna
\mathbf{x}_c &= \left[ \begin{array}{ccccc}
C_{0,0} & C_{10} & C_{11} & \ldots & C_{NN}  \end{array} \right] \label{eq:lsax1}  \ ,
\mathbf{x}_s &= \left[ \begin{array}{ccccc}
S_{11} & S_{21} & S_{22} & \ldots & S_{NN} \end{array} \right] \label{eq:lsax2}  \ .
\ena
We determine the spherical harmonics up to degree $N=100$ and provide them in 
Figure~\ref{fig:spe}. Additional information on the determination of
the gravity field can be found in \cite{Rei15}.

\section{Series in nutation coefficients}


The presence of a large eccentricity of 67P requires to develop 
\equ{e:pre}-\equ{e:serB} up to high orders in $e$: we start 
from \citep[see, e.g.][]{Pet14}:
\bea{nuSER}
&&\Delta\psi=\frac{K}{2}\cos\left(I\right)\times
\int
\left[
\left(\frac{a}{r}\right)^3-\left(\frac{a}{r}\right)^3\cos\left(2\lambda-2h\right)
\right]_{per} dt \ , \nonumber \\
&&\Delta\varepsilon=-\frac{K}{2}\sin\left(I\right)\times
\int
\left[
\left(\frac{a}{r}\right)^3\sin\left(2\lambda-2h\right)
\right]_{per} dt \ .
\eea
Here, $I=-\varepsilon$ is the obliquity angle, $h=-\psi$ is the
precession angle, and  $r$, $\lambda$ are the distance between the Sun and 67P,
and the orbital longitude of the Sun, respectively. Let the angle $\lambda=180^o+
\omega+\nu-\Lambda$, $C'=\cos\left(2\omega-2\Lambda\right)$ and 
$S'=\sin\left(2\omega-2\Lambda\right)$. Since $\omega$, $\Lambda$ are slowly
varying angles with time (in comparison to true anomaly $\nu$) we assume that
$C'$, $S'$ are constant from now on. Using the identities
\bna
&&\cos\left(2\omega+2\nu-2\Lambda-2h\right)=
C'\cos\left(2\nu\right) - S'\sin\left(2\nu\right) \ , \\
&&\sin\left(2\omega+2\nu-2\Lambda-2h\right)=
S'\cos\left(2\nu\right) + C'\sin\left(2\nu\right) \ ,
\ena
we are able to express the integrands in \equ{nuSER}, by means of trigonometric terms
in $2\nu$ instead of $2\lambda-2h$. Making use of basic trigonometric identities, and 
standard series expansions of $\cos\nu$, $\sin\nu$, and 
$\left(a/r\right)^3$ \citep[see, e.g.][]{Stu73} the integrands can also be expressed in 
terms of mean anomaly $M=nt$. The integration with respect to time $t$ provides 
$\Delta\psi$, $\Delta\varepsilon$ in terms of
\bna
\Delta\psi&=&
\sum_k^N\bigg[
c_k(S',n,e)\cos\left(kM\right)+
s_k(C',n,e)\sin\left(kM\right)\bigg] \ , \nonumber \\
\Delta\varepsilon&=&
\sum_k^N\bigg[
c_k'(C',n,e)\cos\left(kM\right)+
s_k'(S',n,e)\sin\left(kM\right)\bigg] \ .
\ena 

We provide the numerical values of the coefficients $c_k$, $s_k$, and $c_k'$,
$s_k'$ of the nutation series for 67P, with $c=0.25$, in Table~\ref{nutA}. The
secular part in the integral of the first equation of \equ{nuSER} (not
depending on mean anomaly $M$) can directly be identified with $\dot\psi$  -
that gives \equ{e:pre}.

\bsp
\label{lastpage}
\end{document}